\documentclass[twocolumn,twoside,slac_two]{revtex4}
\usepackage{graphicx}
\usepackage{fancyhdr}
\newcommand{\G}{\Gamma}

\newcommand{\g}{\gamma}

\newcommand{\psim}{\lower.5ex\hbox{$\; \buildrel \propto \over\sim \;$}}
\newcommand{\lbar}{\lower.0ex\hbox{$\; \buildrel
{\lower0.0ex \hbox{-}} \over\lambda  \;$}}

\newcommand{\tgg}{\tau_{\gamma\gamma}}

\begin{document}

\title{Constraints on the Intergalactic Magnetic Field from Gamma-Ray Observations 
of Blazars}

%

\author{J.\ D.\ Finke}
\affiliation{Space Science Division Code 7653, US Naval Research Laboratory, 4555 Overlook Ave. SW, 
  20375, USA}
\author{L.\ Reyes}
\affiliation{Physics Department, California Polytechnic State University, 
  San Luis Obispo, CA 94307, USA}

\author{M.\ Georganopoulos}
\affiliation{Department of Physics, University of Maryland Baltimore County, 
  1000 Hilltop Circle, Baltimore, MD 21250, USA}
\affiliation{NASA Goddard Space Flight Center, Code 660, Greenbelt, MD 20771, USA}
\author{on behalf of the Fermi-LAT Collaboration}

\begin{abstract}

Gamma rays from distant blazars interact with the extragalactic
background light, creating electron-positron pairs, and reducing the
gamma-ray flux measured by ground-based atmospheric Cherenkov
gamma-ray telescopes.  These pairs can Compton-scatter the cosmic
microwave background, creating a gamma-ray signature observable by the
Fermi Large Area Telesope (LAT).  The signature is also dependent on
the intergalactic magnetic field (IGMF), since it can deflect the
pairs from our line of sight, reducing the gamma-ray emission.  We
present preliminary constraints on the IGMF using Fermi-LAT and
Cherenkov telescope observations, ruling out both very large and very
small values of the IGMF strength.

\end{abstract}

\maketitle

\thispagestyle{fancy}


\section{Introduction}
\label{intro}

The extragalactic backgroung light (EBL) from the infrared (IR),
through the optical and into the ultraviolet (UV) is dominated by
emission, either directly or through dust absorption and reradiation,
of all the stars which have ever existed in the observable universe.
It contains information about the cosmological expansion, star
formation history, dust extinction and radiation in the universe, and
so can potentially provide constraints on a number of cosmologically
interesting parameters.  However, its direct detection is hampered by
the bright foreground emission from the Earth's atmosphere, the solar
system, and the Galaxy.  This can be avoided by measurements from
spacecraft outside the Earth's atmosphere
\citep[e.g.,][]{hauser98,dwek98,bernstein02,mattila03,bernstein07} or
solar system
\citep{toller83,leinert98,edelstein00,murthy01,matsuoka11}, or through
galaxy counts, which in general give lower limits to the EBL
\citep[e.g.,][]{madau00,marsden09}.  See \citet{hauser01} for a review
of EBL measurements, constraints, and models.

Shortly after the discovery of the cosmic microwave background (CMB)
radiation \citep{penzias65}, it was realized that this and other
radiation fields would interact with extragalactic $\g$-rays,
producing $e^+e^-$ pairs, effectively absorbing the $\g$ rays
\citep{gould67_EBL,fazio70}.  In the 1990s, extragalactic high-energy
$\g$-ray astronomy took two major leaps forward, one with the launch
of the {\em Compton Gamma-Ray Observatory} and the first detections of
extragalactic sources at high energies (MeV--GeV) by EGRET
\citep{hartman92}, and another at very-high energies (VHE; $\geq 0.1$\
TeV) by ground-based atmospheric Cherenkov telescopes \citep[ACTs;
e.g., ][]{punch92}.  It was almost immediately realized that $\g$-ray
observations of extragalactic blazars could be used to constrain the
EBL \citep{stecker92,stecker93,dwek94,biller95,madau96}.  The
detection of the hard spectrum from the BL Lac 1ES 1101$-$232 with
HESS seems to put strong constraints on the EBL, ruling out models
that predict high opacity if one assumes the intrinsic VHE spectrum
cannot be harder than $\G=1.5$, where the differential photon flux
$\Phi(E) = dN/dE \propto E^{-\G}$ \citep{aharonian06}.  However, the
$\G=1.5$ constraint has been questioned, and a number of theoretical
possibilities have been raised that could account for harder VHE
spectra \citep*{stecker07_accel,boett08,aharonian08}.  Nonetheless,
the basic idea, that the intrinsic VHE photon index cannot be harder
than a certain number, has been used by many authors to constrain the
EBL
\citep[e.g.,][]{schroedter05_EBL,aharonian07_0229,mazin07,albert08_3c279,finke09_EBLconstrain}.

The new era of $\g$-ray astronomy, began with the launch of the {\em
Fermi Gamma-Ray Space Telescope} has brought additional constraints on
the EBL.  At the energies observed by the LAT, the extragalactic
$\g$-ray sky below $\sim$ 10 GeV is expected to be entirely
transparent to $\g$-rays, at least back to the era of recombination
\citep[$z\sim 1000$; e.g.,][]{oh01,finke10_EBLmodel}, while photons
observed in the 10 GeV -- 300 GeV range should be attenuated by
UV/optical photons if they originate from sources at $z\geq0.5$.  This
suggests a possible way to constrain the EBL.  The LAT spectrum below
10 GeV can be extrapolated to higher energies, and should be an upper
limit on the intrinsic spectrum in the range where the EBL attenuates
the $\g$-rays.  That is, everywhere for the $\g$-ray spectrum
$d^2\log(\Phi)/(d\log(E))^2 \le 0$ (the spectrum is concave).  This
seems to be a reasonable assumption, since no convex $\g$-ray spectrum
has been observed, and it is difficult to imagine theoretical ways to
produce one, although see Section \ref{discussion} for a discussion.
\citet{abdo10_EBL} have used this technique to put upper limits on the
optical/UV EBL absorption optical depth ($\tgg$) for sources $z\geq
0.5$, and rule out with high significance some models which predict
high $\tgg$ with data from the first 11 months of LAT operation from 5
blazars and 2 gamma-ray bursts (GRBs).  More recently,
\citet{ackermann12} used a similar technique in a composite fit to 150
BL Lacs with 46 months of LAT data to constrain the high-$z$ EBL even
further, and found agreement with most recent models
\citep{franceschini08,gilmore09,finke10_EBLmodel,kneiske10,dominguez11,gilmore12_model}.

Below 100 GeV, the universe is expected to be transparent out to
$z\sim 0.1$, although VHE photons from this redshift range should be
attenuated by interactions with IR EBL photons.
\citet*{georgan10_EBL} suggested a very similar technique to
\citet{abdo10_EBL}, applied to the VHE range.  For those sources
detected by both LAT and an ACT, one can extrapolate the LAT spectrum
(which should be unattenuated for these sources at low $z$) into the
VHE range, and use it as an upper limit on the intrinsic VHE flux.  As
with the LAT-only case, a comparison of this upper limit with the
observed VHE flux allows one to compute an upper limit on $\tgg$.
\citet{georgan10_EBL} used this technique to show that models which
predicted high $\tgg$ in the VHE range were strongly disfavored.
Another possible way to estimate the intrinsic spectrum of a source,
and thus estimate $\tgg$, comes from modeling the full radio to GeV
$\g$-ray spectral energy distribution (SED) of a $\g$-ray blazar with
a standard synchrotron/synchrotron self-Compton (SSC) model.  The SSC
spectrum can be extrapolated to the VHE regime, and compared with
observations to estimate $\tgg$ \citep{mank11}.  \citet{dominguez13}
has applied this technique to a sample of $\sim 15$ LAT and
ACT-detected blazars to constrain the cosmic $\g$-ray horizon, i.e.,
the energy where $\tgg=1$ for a certain redshift.  \citet{dwek12}
present a comprehensive review of recent attempts to constrain the EBL
with $\g$-ray observations.

These constraints come with caveats, however.  One possibility is that
the electron positron pairs that are produced by the $\g$-ray--EBL
photon interactions Compton-scatter the CMB photons, producing GeV
$\g$-ray emission which could themselves be absorbed by interactions
with the EBL, producing a cascade
\citep{aharonian94,plaga95,dai02,davezac07}.  If the intergalactic
magnetic field (IGMF) strength is low, the pairs will not be
significantly deflected from our line of sight, and this could produce
an observable feature in the LAT bandpass \citep[e.g.,][]{neronov09}.
Indeed, recently several authors have even used the non-detection of
these cascades to put lower limits on the IGMF strength
\citep[e.g.,][]{neronov10,tavecchio10_igmf,dolag11}.  In general these
efforts depend on the fact that emission from TeV blazars is
relatively constant over long periods of time.  TeV variability has
not been observed in some blazars,
\citep[e.g.,][]{aharonian06,aharonian07_0229}, although many blazars
are highly variable at these energies
\citep[e.g.,][]{aharonian07_2155}.  Observing and studying the
time-dependent EBL-induced pair cascades from GRBs and blazar flares
has been suggested as a way to probe the IGMF parameters
\citep{razzaque04,ichiki08,murase08}.  The possibility of variable TeV
emission has led to caveats in interpreting the IGMF constraints from
apparently non-variable TeV blazars \citep{dermer11,taylor11}.  Other
uncertainties such as the EBL intensity and TeV spectra errors, blazar
jet geometry and Doppler factor, can further decrease the lower limits
on the IGMF \citep{arlen12}.

In this paper we report on our efforts to constrain both the EBL
absorption optical depth, $\tau_{\g\g}$ and the IGMF strength
($B_{IG}$) and coherence length ($\lambda_{B}$) using $\g$-ray
observations of the blazars 1ES~1101$-$232 and 1ES~0229+200 from both
LAT and ACTs.  This technique can be seen as an extension of previous
work by \citet*{georgan10_EBL}.  We make use of data from the first
3.5 years of LAT operation, the analysis of which is described in
Section \ref{lat_analysis}, and ACT spectra from the literature.  Our
technique for constraining the IGMF is described in Section
\ref{method}.  We report our preliminary constrains on the IGMF
parameters in Section \ref{constraintIGMF}.  Finally we discuss these
results, their caveats and implications in Section \ref{discussion}.

\section{LAT Analysis}
\label{lat_analysis}

To determine the LAT spectra of 1ES~1101$-$232 and 1ES 0229+200 we
considered all LAT events collected since the start of the mission for
3.5 years of operation (i.e. until 2012 February 16), a significant
increase in statistics with respect to previous efforts. The data were
analyzed using an official release of the {\em Fermi} ScienceTools (v9r27)
and Pass 7 instrument response functions. Only photons satisfying the
CLEAN event selection are included in the analysis in order to reduce
the likelihood that charged-particle background particles are mistaken
as high-energy photons. As usual, the spectral analysis of each source
is based on the maximum likelihood technique using the standard
likelihood analysis software. In both cases we considered all the
point-like sources within $15^\circ$ from the source position (as
determined from the 2FGL catalog and improved diffuse background
components (Galactic and extragalactic). A likelihood ratio test was
used to find the best spectral model (power-law, broken power law, and
log parabola) that fits the data. In both cases, there was
insufficient evidence for deviation from a power-law spectrum.

In addition to the standard maximum likelihood analysis performed to
find the best fit to the data, the Log-of-the-likelihood ($LL$)
profile as a function of the source's flux normalization ($F_0$) was
calculated in order to fully characterize the uncertainty on this
parameter. Thus, instead of assuming a perfect Gausian distribution
for the error of $F_0$ and using 1, 2, and 3 standard deviations to
calculate the 68\%, 95\% and 99\% confidence intervals, we used the
$LL$ profile to calculate the actual confidence intervals assuming
that $-2\Delta(LL)$ is distributed as the chi-square probability
distribution with one degree of freedom. We found that this approach
is necessary in order to correctly determine the flux probability
distribution of weak LAT sources such as 1ES~0229+200 and 1ES
1101$-$232.

\section{Method for Constraining Models}
\label{method}

Our technique for constraining the EBL and IGMF is based on
\citet{georgan10_EBL}, with extentions including a sophisticated Monte
Carlo (MC) technique to accurately determine the significance of the
constraints, and the addition of a pair cascade component.

\subsection{Technique Assumptions}
\label{assumptions}

We make the following assumptions, the first three of which are idential 
to the ones of \citet{georgan10_EBL}:

{\em \#1} We assume the MeV-TeV flux from BL Lacs in our sample are
produced cospatially from the sources themselves (and not UHECR interactions 
in intergalactic space), and that they are never convex (i.e., never is
$d^2\log(\Phi)/(d\log(E))^2 > 0$). 
 
{\em \#2} We assume the objects are not variable at $\g$-rays within the
statistical uncertainties of the measurements.  Indeed, we have
selected sources for our sample for which $\g$-ray variability has not
been reported, in either the LAT or ACTs.

{\em \#3} We assume the $\g$ rays will not avoid being absorbed on their
path to Earth, by converting to axions or some other exotic mechanism
\citep[e.g.,][]{deang07,sanchez09}

{\em \#4} We assume pairs created by $\g$-ray-EBL photon
interactions will Compton-scatter the CMB, and will lose energy
primarily through scattering and not through intergalactic plasma beam
instabilities
\citep{broderick12,schlickeiser12_pair,schlickeiser12_plasma}.  Our
technique for including the cascade component is similar to the one
used by \citet{meyer12_EBL}.

We discuss the viability of these assumptions in Section
\ref{discussion}.

\subsection{Ruling out Models}
\label{ruleout}

\begin{figure}[t]
\centering
\vspace{2.2mm} 
\includegraphics[width=70mm]{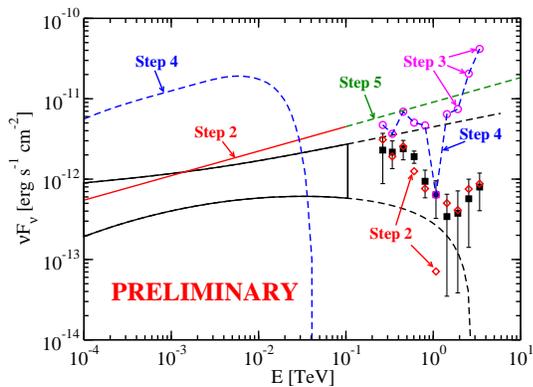}
\caption{ This figure illustrates many of the steps in our method for
ruling out models from Section \ref{ruleout}, using the $\g$-ray
spectrum for 1ES 1101$-$232.  The LAT spectrum is shown as the bowtie,
along with this spectrum extrapolated to the VHE regime as the dashed
curves.  The observed HESS spectrum is shown as the filled
squares. The randomly drawn HESS points shown as empty diamonds
($F_{VHE}$) and the randomly drawn LAT spectrum is shown as a line,
both of which are labeled ``Step 2''.  The deabsorbed points are shown
as the circles ($F_{VHE,int}$) and labeled ``Step 3''.  The cascaded
component and the interpolated VHE spectrum used to calculate it are
shown as dashed curves labeled ``Step 4''. The LAT spectrum
extrapolated into the VHE regime is shown as the dashed line labeled
``Step 5''.  For the MC iteration shown here, the model is ruled out
by both criteria in Step 5, since $F_{LAT} < F_{cascade}$ and for
several points $F_{LAT,ext} < F_{VHE,int}$.  }
\label{methodfig}
\vspace{2.2mm}
\end{figure}

Our technique for ruling out a model is illustrated in Figure \ref{methodfig} and 
has the following steps:

{\em Step 1.} Select the model we wish to rule out.  This includes selecting
an EBL model (taken from the literature), selecting an IGMF strength
($B_{IG}$) and coherence length ($\lambda_B$); and selecting a blazar
jet radiation opening angle and a ``blazar lifetime'' ($t_{blazar}$),
that is, the length of time the blazar has been emitting $\g$-rays
with its current luminosity \citep{dermer11}.

{\em Step 2.} Given the integrated LAT flux ($F_{LAT}$) and photon
index ($\G$) and their errors for a particular blazar (see Section
\ref{lat_analysis}), we draw a random $F_{LAT}$ and $\G$ from a
probability distribution function (PDF) which represents their errors.

{\em Step 3.} For each energy bin of the VHE spectrum, with a measured flux
and error, we draw a random flux, $F_{VHE}$, assuming the flux errors
are distributed as a normal distribution.  Each randomly drawn
$F_{VHE}(E)$ is deabsorbed with the EBL model we are testing to give
an intrinsic flux, $F_{VHE,int}(E) = \exp(\tau_{\g\g}(E))\times
F_{VHE}(E)$.

{\em Step 4.} From $F_{VHE,int}(E)$, the contribution of the $e^+ e^-$ pairs
Compton-scattering the CMB, $F_{cascade}$ is calculated.  The
deabsorbed emission is assumed not to extend beyond the highest energy
bin, the most conservative assumption we can make.  The cascade flux
$F_{cascade}$ is calculated using the formula of \citet{dermer11} and
\citet{dermer12}.  The minor corrections to this formula from
\citet{meyer12_EBL} should have no real effect on the results, since
they affect the lowest flux portion of the cascade spectrum.

{\em Step 5.} The randomly drawn LAT power-law spectrum from step 2 is
extrapolated to the VHE regime.  For this MC iteration, the model is
considered rejected if one of two criteria are met: (i) the cascade
flux $F_{cascade}$ exceeds the randomly drawn LAT integrated flux
$F_{LAT}$; or (ii) any one of the deabsorbed flux bins from
$F_{VHE,int}(E)$ exceed the extrapolated LAT flux, $F_{LAT,ext}$, {\em
unless} $0.01 F_{LAT} \le F_{cascade} < F_{LAT}$, in which case the
model is never rejected for this iteration.  If the cascade flux makes
up a significant fraction of the observed LAT flux, we do not believe
it can be extrapolated to the VHE regime and used to constrain models.
The rejection criteria in this step are based on ones from
\citet{georgan10_EBL} and \citet{meyer12_EBL}.

{\em Step 6.} Steps 2--5 above are repeated $N_{trials}$ times (we use
$N_{trials}=10^6 $) and the number of times the model is rejected
$N_{reject}$ is counted.  The probability the model is ruled out is
$P_{reject}=N_{reject}/N_{trials}$.

\section{Constraints on IGMF}
\label{constraintIGMF}

We apply the method outlined in Section \ref{ruleout}, to $\gamma$-ray
observations of 1ES~0229+200 and 1ES~1101$-$232.  We use the HESS
spectra for these sources from \citet{aharonian07_0229} and
\citet{aharonian07_1101}, respectively, and the LAT spectra calculated
as outlined in Section \ref{lat_analysis}.  We used the EBL model from
\citet{finke10_EBLmodel}, a jet radiation opening angle of 0.1 rad,
and a blazar lifetime of $t_{blazar}=H_0^{-1}$, where $H_0$ is the
Hubble constant; that is, we assume the blazar has been emitting for
essentially the entire age of the universe.  Neither of these objects
have been shown to be variable at $\g$-ray energies, with either the
LAT or TeV instruments.

\begin{figure}[t]
\centering
\vspace{2.2mm} 
\includegraphics[width=80mm]{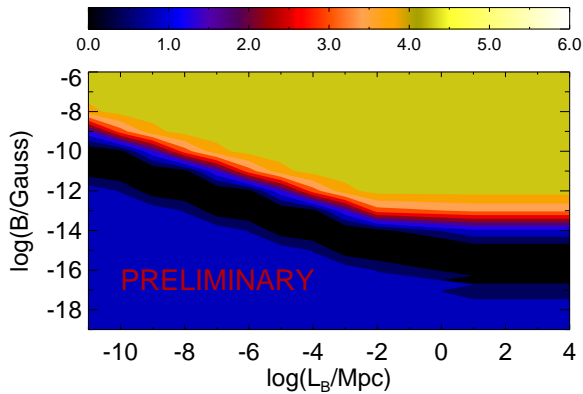}
\caption{ The values of parameter space of $B$ and $L_B$ which are
ruled out by our method, using LAT and HESS observations of
1ES~0229+200.  The significance that the values are ruled out are
given by the colors, in number of sigma, as indicated by the color
bar.  }
\label{contour_0229}
\vspace{2.2mm}
\end{figure}

\begin{figure}[t]
\centering
\vspace{2.2mm} 
\includegraphics[width=80mm]{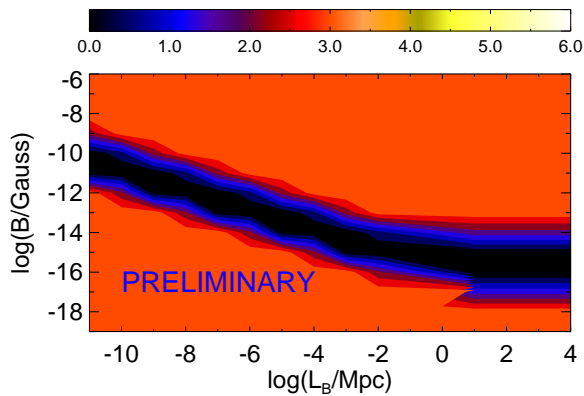}
\caption{ Similar to Fig.\ \ref{contour_0229}, only for 1ES~1101$-$232.}
\label{contour_1101}
\vspace{2.2mm}
\end{figure}

The results from the application to 1ES~0229+200 and 1ES~1101$-$232 can
be found in Figs. \ref{contour_0229} and \ref{contour_1101},
respectively.  The probabilities that values of parameter space are
ruled are converted to number of sigma, assuming the errors are
distributed normally.  High values of $B_{IG}$ are ruled out by the
constraint that deabsorbed TeV flux cannot exceed the LAT spectrum
spectrum extrapolated into this regime; and low values of $B$ are
ruled out by the constraint that the cascade cannot exceed LAT flux
(see Section \ref{ruleout}, step 5).  A black strip in these figures
indicates allowed values of $B_{IG}$ and $L_B$.  Here the cascade flux
was less than the observed LAT flux, but $>0.01$ of the LAT flux, so
that the LAT flux could not be simply extrapolated into the TeV
regime.

\begin{figure}[t]
\centering
\vspace{2.2mm} 
\includegraphics[width=80mm]{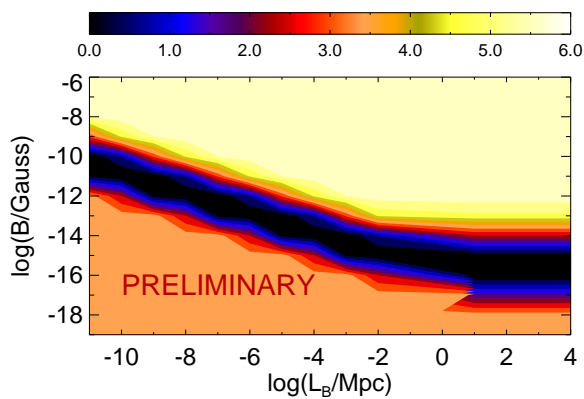}
\caption{ Similar to Figs.\ \ref{contour_0229} and \ref{contour_1101}, 
only here we show the combined results for both 1ES~1101$-$232 and 1ES~0229+200.}
\label{contour_tot}
\vspace{2.2mm}
\end{figure}

The combined results for both 1ES~0229+200 and 1ES~1101$-$232 are shown
in Fig. \ref{contour_tot}.  High values of $B_{IG}$ are ruled out at
greater than $5\sigma$, while lower values are ruled out at greater
than $3\sigma$.

\section{Discussion}
\label{discussion}

We have used LAT and TeV data to constrain IGMF parameters.  Our
preliminary results indicate high values of $B_{IG}$ are ruled out
with $>5\sigma$ significance.  This means that either one of our model
assumptions is not correct (Section \ref{assumptions}), one of our
parameters is not correct, or the cascade flux {\em must} contribute
at least 1\% to the emission observed by the LAT.  We discuss these
possibilities below.

If the GeV and TeV $\g$-rays are not produced co-spatially (model
assumption \#1), this could explain our results at high $B_{IG}$.  It
is possible they are not
\citep{boett08,essey10_1,essey10_2,essey11_cr}.  In this case one
might expect the TeV flux to be larger than the extrapolated LAT flux.
If the two sources we used are variable (assumption \#2), the
variablity would have to be greater than the observational errors,
which are quite large, to invalidate our results.  Our results could
be interpreted as evidence for axion-like particles (assumption \#3),
which would allow TeV photons to avoid much of the EBL attenuation.
This would be difficult to distinguish, although with sufficiently
accurate observations certain $\g$-ray signatures may be found
\citep[e.g.,]{sanchez09}.  Finally, if all our other assumptions are
correct, the cascade component is likely to exist at the $>5\sigma$
level.  {\em This implies that it is very unlikely that plasma beam
instabilities would eliminate the cascade} \citep[][assumption
\#4]{broderick12}.  See also \citet{venters12} for a critical
assesment of this possibility.

Of our model parameters, the EBL model of \citet{finke10_EBLmodel} is 
unlikely to be wrong to a large degree, given its proximity to 
galaxy counts and agreement with other models.  The timescale 
the blazar has been operating would not effect the upper limits on 
$B_{IG}$, nor could the jet opening angle.  We will explore other 
EBL models in future work.

The final possibility is that there is indeed a significant cascade
component created, and the upper limits on $B_{IG}$ and $L_{B}$ can be
taken at face value.  We note that in this case, the parameters are in
the range where one might expect extended GeV emission (``$\g$-ray
halos'') around TeV sources to be detectable by the LAT
\citep{neronov09}.  The detection of this emission or lack thereof
could provide confirmation of these parameter constraints, or the
other possibilities discussed above.  The detection of these halos in
the {\em Fermi} era has been controversial
\citep{ando10_halo,neronov11,ackermann13_psf}.

The constraints on lower magnetic field values from the cascade are
less certain, not just because of their lower significance but also
due to their dependence on parameter assumptions.  The timescale the
blazar has been emitting TeV $\g$-rays or changes the opening angle
can significantly effect the cascade \citep{dermer11,arlen12}.  We
will explore how changes of these parameters effect our results.

The IGMF may have been generated in phase transitions in the early
universe \citep[e.g.,][]{grasso01}.  Our {\em preliminary} results
ruling out low $B_{IG}$ disfavor IGMFs generated from electro-weak
phase transitions, if the generated fields are non-helical
\citep{neronov09}.  However, as discussed above, the lower magnetic
field values are dependent on caveats including those regarding source
variability.  If the IGMF is generated from phase transitions, and if
it is possible to constrain $B_{IG} \ge 10^{-15}$\ G, then it will be
almost impossible to detect gravitational waves from inflation
\citep{fujita12}.  The IGMF parameters also have implications for the
diffuse gamma ray background and its anisotropy
\citep{venters12,ackermann12_diffuse_aniso}.

\begin{acknowledgments}

The {\em Fermi} LAT Collaboration acknowledges support from a number of
agencies and institutes for both development and the operation of the
LAT as well as scientific data analysis. These include NASA and DOE in
the United States, CEA/Irfu and IN2P3/CNRS in France, ASI and INFN in
Italy, MEXT, KEK, and JAXA in Japan, and the K.~A.~Wallenberg
Foundation, the Swedish Research Council and the National Space Board
in Sweden. Additional support from INAF in Italy and CNES in France
for science analysis during the operations phase is also gratefully
acknowledged.

\end{acknowledgments}

\bigskip 

\bibliography{3c454.3_ref,EBL_ref,references,mypapers_ref,blazar_ref,sequence_ref,SSC_ref,LAT_ref}





\end{document}